\documentclass[fleqn,usenatbib, letters]{mnras}

\usepackage{savesym}
\usepackage{amsmath}
\savesymbol{iint}
\usepackage{txfonts}
\restoresymbol{TXF}{iint}

\usepackage[T1]{fontenc}
\usepackage{ae,aecompl}

\usepackage{graphicx}	
\usepackage{amssymb}	

\title[Magnetic fields in protoplanetary disks]{Magnetic fields in protoplanetary disks: from MHD simulations to ALMA observations}

\author[G. H.-M. Bertrang et al.]{
G. H.-M. Bertrang,$^{1,2,3,4}$\thanks{E-mail: bertrang@das.uchile.cl}
M. Flock,$^{5}$
S. Wolf$^{1}$
\\
$^{1}$Kiel University, Institute of Theoretical Physics and Astrophysics, Leibnizstr. 15, 24118 Kiel, Germany\\
$^{2}$Universidad de Chile, Departamento de Astronom\'ia, Casilla 36-D, Santiago, Chile\\
$^{3}$Universidad Diego Portales, Facultad de Ingenier\'ia, Av. Ej\'ercito 441, Santiago, Chile\\
$^{4}$Millennium Nucleus Protoplanetary Disks in ALMA Early Science, Universidad de Chile, Casilla 36-D, Santiago, Chile\\
$^{5}$Jet Propulsion Laboratory, California Institute of Technology, Pasadena, California 91109, USA
}

\date{Accepted 2016 September 7. Received 2016 September 7; in original form 2016 July 22}

\pubyear{2016}

\AtBeginShipout{%
  \ifnum\value{page}>1 %
    \typeout{* Additional boxing of page `\thepage'}%
    \setbox\AtBeginShipoutBox=\hbox{\copy\AtBeginShipoutBox}%
  \fi
}

\begin{document}
\label{firstpage}
\pagerange{\pageref{firstpage}--\pageref{lastpage}}
\maketitle

\begin{abstract}
Magnetic fields significantly influence the evolution of protoplanetary disks and the formation of planets, following the predictions of numerous magnetohydrodynamic (MHD) simulations. However, these predictions are yet observationally unconstrained. To validate the predictions on the influence of magnetic fields on protoplanetary disks, we apply 3D~radiative transfer simulations of the polarized emission of aligned aspherical dust grains that directly link 3D global non-ideal MHD simulations to ALMA observations.  Our simulations show that it is feasible to observe the predicted toroidal large-scale magnetic field structures, not only in the ideal observations but also with high-angular resolution ALMA observations. Our results show further that high angular resolution observations by ALMA are able to identify vortices embedded in outer magnetized disk regions. 
\end{abstract}




\section{Introduction}
Magnetic fields play a crucial role in the formation and evolution of protoplanetary disks \citep{2011ARA&A..49...67W, 2011ARA&A..49..195A}.  Magnetic fields influence the transport of dust and gas 
\citep[e.g.,][]{2007ApJ...654L.159C, 2014prpl.conf..411T, 2014prpl.conf..339T}, the disk
chemistry \citep[e.g.,][]{2011ApJS..196...25S, 2013ChRv..113.9016H}, and even the migration of 
plane\-tesimals and planets within the disk \citep[e.g.,][]{2010A&A...515A..70D, 2011MNRAS.415.3291G, 2011ApJ...736...85U} through 
magnetohydrodynamic (MHD) turbulence. Moreover, MHD turbulence can provide the source of viscosity that drives the accretion \citep{1974MNRAS.168..603L}, and thus, the evolution of the disk \citep{1973A&A....24..337S}.  One of the most promising
mechanisms for driving turbulence, respectively accretion is the 
magneto-rotational instability \citep[MRI;][]{1991ApJ...376..214B, 1996ApJ...467...76B, 1998RvMP...70....1B}.  
 Turbulence generated by pure hydrodynamical instabilities \citep{2013MNRAS.435.2610N, 2014ApJ...789...77L, 2014ApJ...788...21K} is not able
to explain the observed accretion rates and disk lifetimes \citep{2011ARA&A..49...67W}, despite such instabilities could become important in disk regions with low ionization \citep{2014prpl.conf..411T}. High enough ionization, and so magnetic activity, is expected in the inner disk regions due to thermal ionization \citep{ume88,des15} 
and also in the outer disk regions due to interstellar radiation \citep[e.g.,][]{2007ApJ...659..729T,2013ApJ...765..114D, 2015ApJ...799..204C, 2015A&A...574A..68F}. However, even the prediction of any magnetic field structure by observational constraints remains a difficult task. First polarimetric observations of protoplanetary disks performed with SMA, CARMA, and VLA indicate toroidal magnetic field structures \citep{2015ApJ...814L..28C, 2014ApJ...780L...6R, 2013ApJ...769L..15S}. Spatially resolved observations of polarized millimeter continuum emission of aligned aspherical dust grains are well suited to reveal the magnetic field structure in the protoplanetary disk \citep[e.g.,][]{2000prpl.conf..247W, 2007ApJ...669.1085C}. Though, polarimetry is strongly influenced by many factors \citep{Bertrang2016}. Taking the dust grain shape, dust grain alignment, magnetic 
field properties, the resolution of the observation, and the projection along the line of sight into account, we present a feasibility study on high-angular resolution polarimetric observations and their interpretation to validate MHD predictions on magnetic fields in protoplanetary disks.

\section{Methods and model}

\subsection{Framework}
 
For this work, we combine state-of-the-art 3D non-ideal MHD simulations with a new Monte-Carlo radiative transfer algorithm to calculate for the dust polarization emission. The resulting synthetic images are then post-processed with the CASA software package (v4.5.2) to obtain realistic ALMA maps.

The 3D global non-ideal stratified MHD simulations were performed using the FARGO MHD code PLUTO \citep{2012A&A...545A.152M, 2015A&A...574A..68F}.

The radiative transfer simulations are performed with an extended version of the code MC3D
\citep{Bertrang2016,1999A&A...349..839W, 2003CoPhC.150...99W}. We use the MHD models as input for the dust density distribution and magnetic field topology, and compute the temperature distribution and anisotropy of the radiation field self-consistently. We assume aspherical dust grains with an axis ratio of $1:1.3$. We further assume here that the grains are perfectly aligned by the magnetic field. Based on these assumptions, we compute an upper limit for the polarized dust emission for aspherical grains. 

\begin{table}
 \caption{Model parameters}
 \label{tab:model}
 \begin{tabular}{ l c }\\\hline\hline
   Parameter & Value\\\hline
                     & Stellar properties \\\hline
Distance (pc) & 100 \\
Effective temperature (K) & 4000 \\
Luminosity (L$_{\odot}$) & 0.95 \\
Mass (M$_{\odot}$) & 0.5 \\\hline
                & Disk parameters \\\hline
Total mass (M$_{\odot}$) & 0.085 \\
Gas-to-Dust ratio &  100 \\
Disk inclination &  0$^\circ$, 45$^\circ$, 90$^\circ$ \\\hline
                & Dust parameters \\\hline
Grain size distr. ($\mu$m)  &   $5\cdot10^{-2}-1\cdot10^{4}$\\
Grain size distr. exponent & -3.5 \\
Grain Composition &  Si ($62.5\%$), C ($27.5\%$) \\
Axis ratio & 1.0:1.3\\\hline
                & Grid parameters \\\hline
$N_r \times N_{\theta} \times N_{\varphi}$  &  256 $\times$ 128 $\times$ 512 \\
$\Delta r_{au} \times \Delta \theta_{rad} \times \Delta \varphi_{rad}$ & 20-100 $\times$ 0.72 $\times$ 2$\pi$ \\
  \hline\hline
 \end{tabular}
\end{table}

Finally, to compare our MHD simulations with ALMA observations, we apply the CASA software package (v4.5.2). We si\-mu\-late observations of 1.5h~integration time at Band~6 (7.5GHz bandwidth) with the antennae configuration C40-5 and a spatial resolution of 0.16". ALMA  is able to detect the degree of polarization for resolved sources down to 0.3\%, along with an uncertanty of 6$^{\circ}$ in the polarization angle \citep{ALMAMan4}.

\subsection{Disk model}
The disk model for the global 3D non-ideal MHD simulations  
follows an initial gas surface density of $\Sigma = 5.94 g\, cm^{-2} \left ( \frac{100 au} {R} \right )$,
with the cylindrical radius R. Temperature and density profiles
are in fully radiation hydrostatic equilibrium. The simulation domain in spherical coordinates spans from $20$au to $100$au in radius, $\Delta \theta = 0.72$ rad in $\theta$ and full $2 \pi$ in azimuth. We include an initial vertical magnetic field with an 1/R profile in radius. The initial resistivity profile is calculated using the dust chemistry method by
\citet{2013ApJ...765..114D}, including HCO+ as dominant ion, electrons, and charged dust. For more details we refer to model D2G\_e-2 by \citet{2015A&A...574A..68F} (Section 2 and Fig. 1 therein). 

For the radiative transfer simulations, we take two typical states of the global MHD simulations as input for the dust density structure and the magnetic field topology. The ring state, a state in which a axisymmetric gap and jump
structure is present, and the vortex state, in which the ring shows a vortex and an enhanced density clump. Both states are emerging at the dead-zone edge and for more details on the setup and models we refer the reader to our previous works \citep{2015A&A...574A..68F,2016A&A...590A..17R}. 
The model parameters are summarized in Tab.~\ref{tab:model}.

\section{Results}
We aim at exploring the feasibility of high-angular resolution polarimetric observations and their interpretation to validate MHD predictions on magnetic fields in protoplanetary disks \citep[for a discussion of the unpolarized dust emission, see][]{2015A&A...574A..68F,2016A&A...590A..17R}. The questions arising from our MHD results are: Can these magnetic field structures actually be traced? How can the differences between these models be detected to determine the physics taking place in a certain observational object? To answer these questions, we focus on simulated $1.3$mm maps of our models. We assume a distance to the disk of $100$pc. 

\begin{figure*}
\includegraphics[width=0.3\textwidth, keepaspectratio]{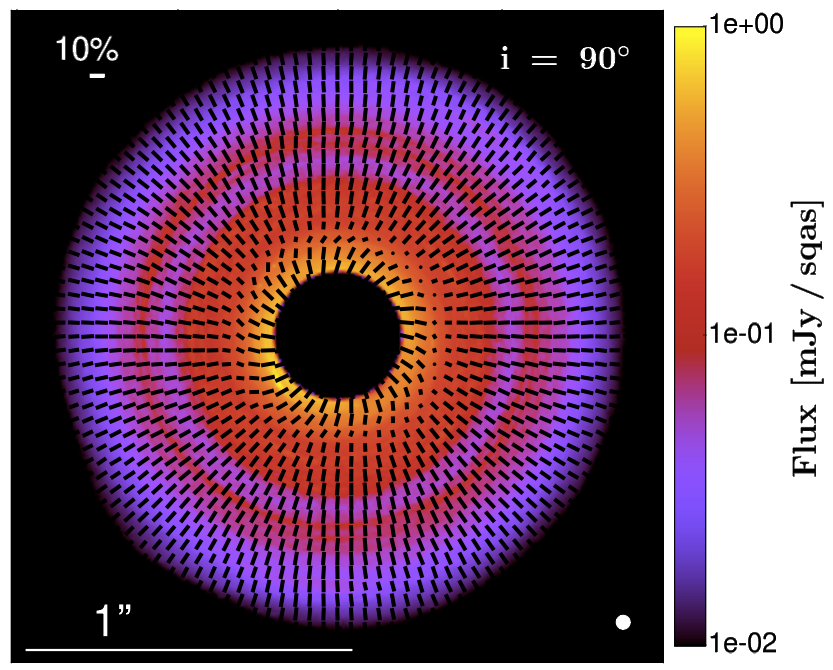}
\includegraphics[width=0.3\textwidth, keepaspectratio]{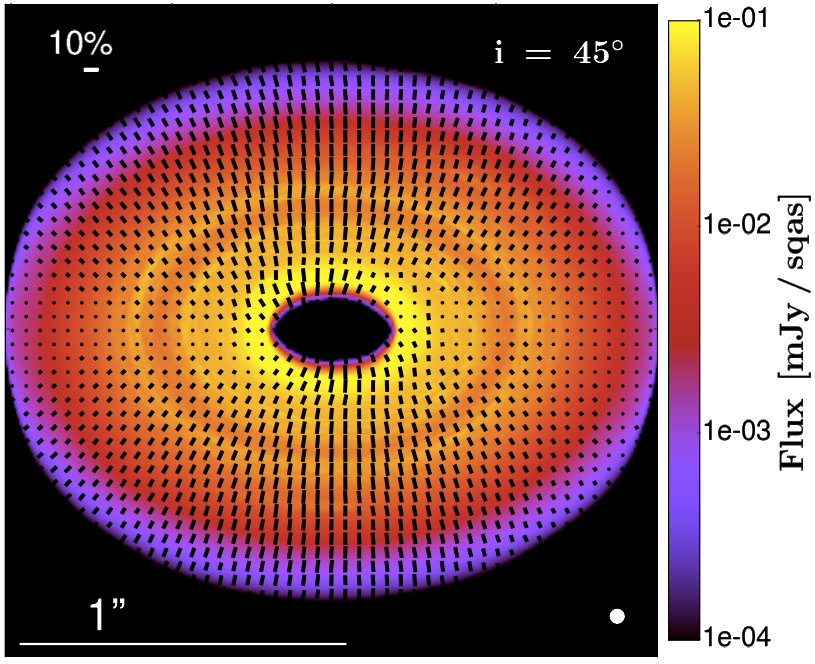}
\includegraphics[width=0.3\textwidth, keepaspectratio]{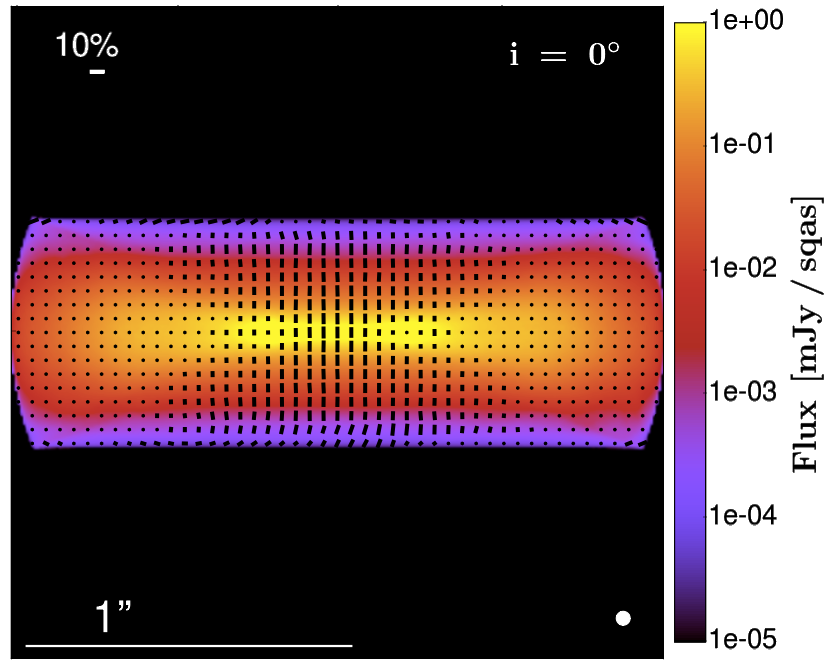}\\
\includegraphics[width=0.3\textwidth, keepaspectratio]{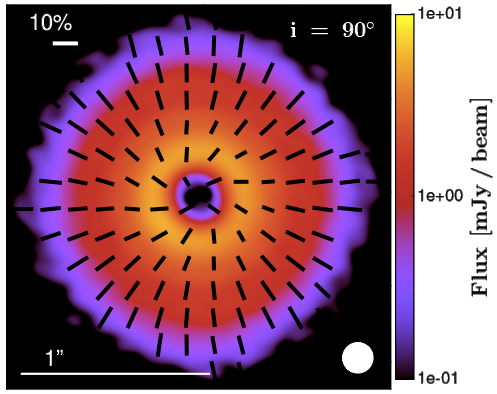}
\includegraphics[width=0.3\textwidth, keepaspectratio]{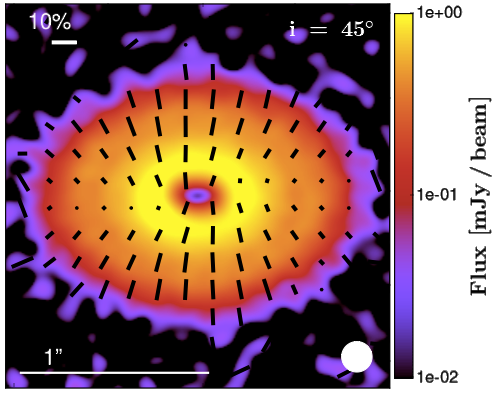}
\includegraphics[width=0.3\textwidth, keepaspectratio]{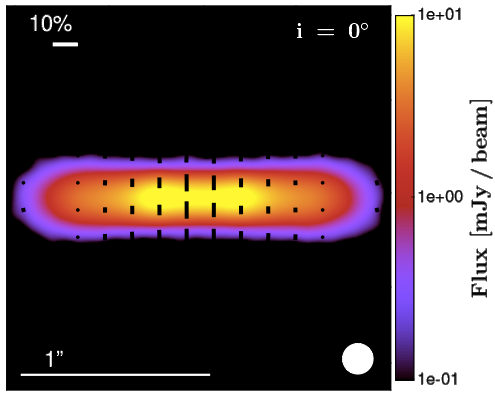}
\caption{The ring state: Self-consistent radiative transfer simulations of the polarized emission of aligned aspherical dust grains at $1.3\,$mm ({\it top row}), and simulated ALMA observations ({\it bottom row}; CASA v5.4.2, band~6, configuration C40-5, $1.5$h integration time; minimal displayed polarization degree is 1\%, i.e., the 3$\sigma$ level of ALMA) at three different disk orientations ({\it left:} face-on, {\it center:} $45^{\circ}$, {\it right} edge-on). The color map shows the total (unpolarized) intensity which is overplotted by polarization vectors.The vectors are plotted with the spatial resolution indicated by the beam size given as white ellipses.  The toroidal magnetic field topology is traced by its characteristic polarization structure in all of the three orientations in the simulated ALMA observation.}
\label{pic:T3maps}
\end{figure*}

\begin{figure*}
\includegraphics[width=0.3\textwidth, keepaspectratio]{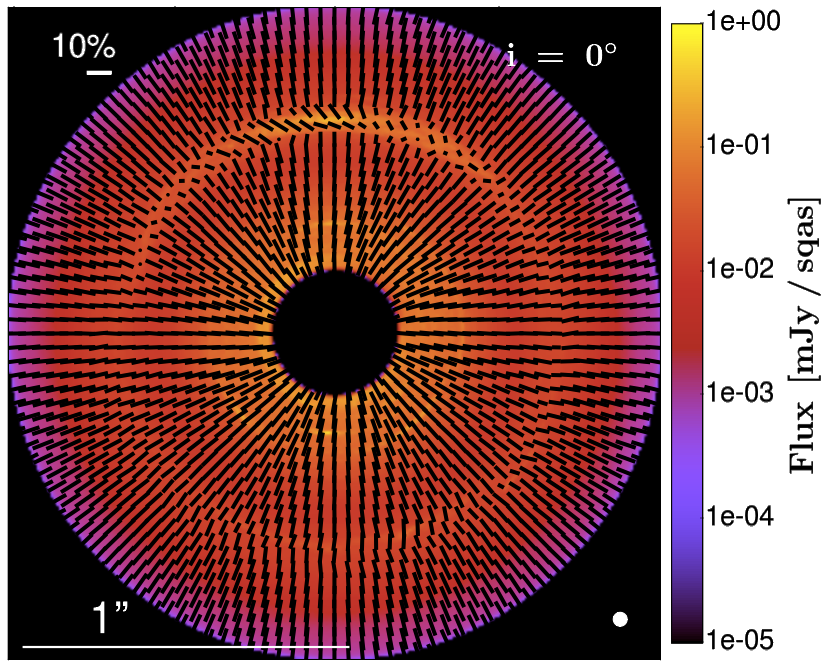}
\includegraphics[width=0.3\textwidth, keepaspectratio]{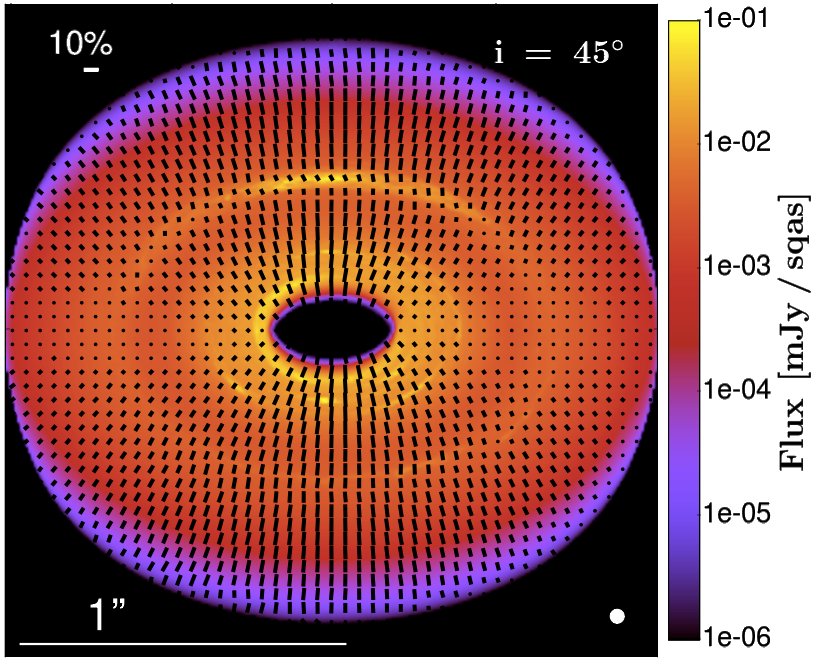}
\includegraphics[width=0.3\textwidth, keepaspectratio]{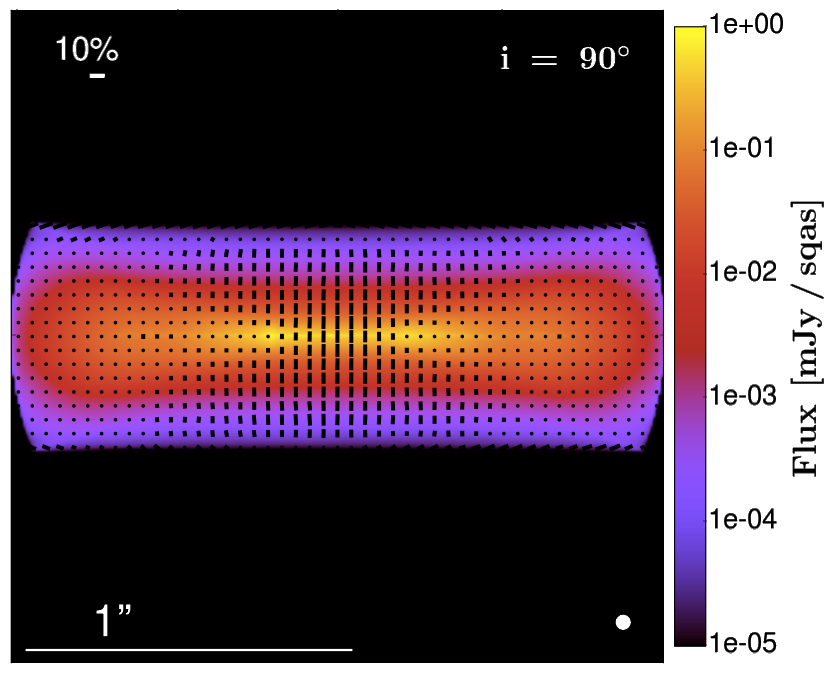}\\
\includegraphics[width=0.3\textwidth, keepaspectratio]{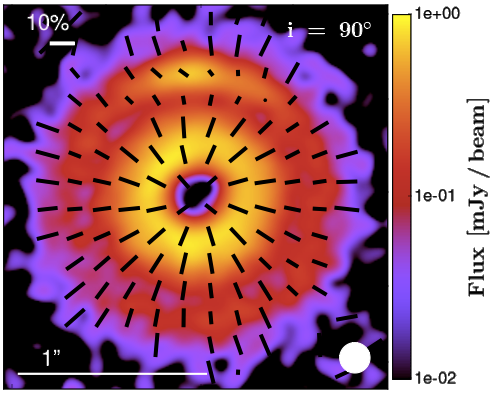}
\includegraphics[width=0.3\textwidth, keepaspectratio]{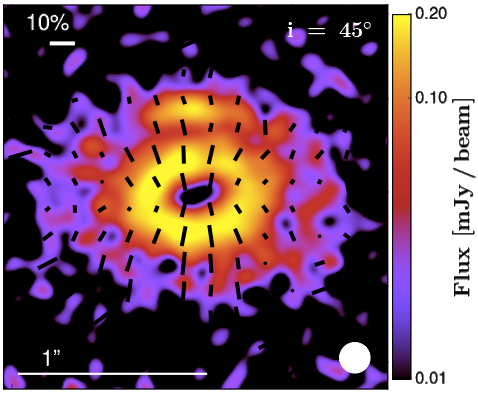}
\includegraphics[width=0.3\textwidth, keepaspectratio]{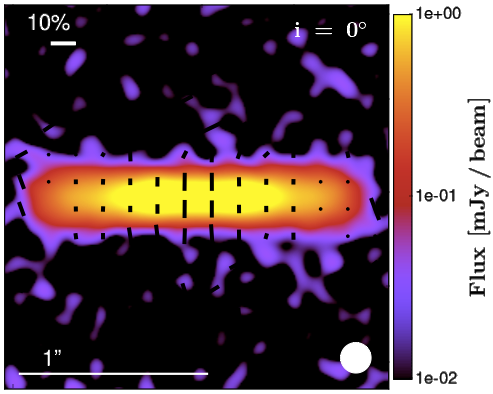}
\caption{The vortex state: As in Fig.~\ref{pic:T3maps}.  The toroidal magnetic field topology is traced by its characteristic polarization structure in both orientations in the simulated ALMA observation, while the vortex is clearly detected in the simulated ALMA observation of a face-on disk.}
\label{pic:T8maps}
\end{figure*}

In Figures~\ref{pic:T3maps}~and~\ref{pic:T8maps}, we present the post-processing radiative transfer simulations and simulated ALMA observations of the two typical states of the global MHD simulations at three different orientations of the disk.  We find that the magnetic field topology as it is predicted by MHD simulations, can be validated by ALMA. 

We find that the polarized signal can be traced across the extend of the disk, in both the ideal observation (the direct result of the radiative transfer simulation) as well as the simulated ALMA observation. The local degree of polarization lies between $1\% - 10\%$ (on a 3$\sigma$ level) for the given model properties. While the degree of polarization in the ideal observation varies with the underlying density structure and disk inclination, it changes additionally with the noise level (i.e., the detection limit) and spatial resolution (the lower the spatial resolution, the greater the impact of signal annihilation due to different polarization angles; e.g., an unresolved radial symmetric polarization pattern results in zero polarization) of the simulated interferometric measurement. This is caused by the sensitivity of the degree of polarization for projectional effects. 

Further, we find that the polarized signal of a face-on disk shows a large-scale radial symmetric pattern, again in both the simulated ideal observation as well as the ALMA observation, for both models.  Our MHD simulations show that the toroidal magnetic field topology continues within the density gaps \citep{2015A&A...574A..68F}, and accordingly, we find that the radial polarization pattern crosses these regions. Thus, the dust density gaps are traceable only by the unpolarized emission \citep[and applying a higher spatial resolution, see][]{2015A&A...574A..68F}.
However, we find clear deviations from the dominant toroidal structure inside the density enhancent region of the vortex which can be traced in the simulated face-on configuration (see Fig~\ref{pic:T8maps}, left).

At an inclination of 45$^{\circ}$, we find a change in the radial polarization pattern towards a more homogenous, parallel structure. Towards the wings of the disk, the degree of polarization nulls. In the simulated ALMA observation, we find that the vortex can only be found in the total (unpolarized) intensity due to the density enhancement inside it. In comparison with the polarization signal in the ideal observation arising from the vortex at a face-on position, the polarization signal at an inclination of 45$^{\circ}$ appears to be altered by the vortex much less strongly. The spatial extent of the vortex in the polarization pattern appears to be reduced significantly (in the order of a factor of 2), and the change in the polarization angle and degree is accordingly smaller (see below). These differences are not traced anymore in the simulated ALMA observation. Thus, the polarization patterns of the two typical states of the MHD simulations  at this inclination are only distinguishable in the ideal observation, not in the simulated ALMA maps.

Additionally, we find for an edge-on oriented disk that the resulting polarization pattern is homogenous and parallel in its orientation. The degree of polarization is maximal along the vertical center axis of the disk and drops quickly towards the wings of the disk. This behaviour is typical for an underlying toroidal magnetic field structure and is also traced by the simulated ALMA observation. These findings as well as those of less inclined disks correspond to observations of the polarization pattern of younger (class~I) protoplanetary disks that have been performed so far \citep{2013ApJ...769L..15S, 2014Natur.514..597S}, whilst the comparison to class~II disks continues to wait for corresponding detections.

On the small spatial scales, we find that the polarized signal traces vortices in the magnetic field structure in both the degree of polarization and the orientation of the polarization pattern in the case of a face-on disk. Vertical components of the magnetic field as well as unordered field orientations in vortices, as resulting from our MHD simulations, act depolarizing due to projectional effects along the line-of-sight which emphasizes the importance of a full 3D~disk model. Along the more ordered magnetic field lines found in large-scale vortices such as in Fig.~\ref{pic:T8maps} the orientation of the polarization pattern follows the vortex geometry. The vortex in the magnetic field structure alters the polarization pattern with respect to a purely radial structure locally by up to 90$^{\circ}$ in the polarization angle, and depolarizes the signal locally down to anhiliation (Fig.~\ref{pic:T8maps}). Our simulations show that these effects cannot only been seen in the ideal observations but also with high-angular resolution ALMA observations. The influence of vertical magnetic field components, however, is not traceable with the presented resolutions in the case of a more inclined disk.

\section{Summary and Conclusion}
We present a feasibility study on high-angular resolution polarimetric observations and their interpretation to validate MHD predictions on magnetic fields in protoplanetary disks.  Based on the assumption of perfect grain alignment by the magnetic field, we compute an upper limit for the polarized dust emission for aspherical grains. This is consistent with previous work \citep{2007ApJ...669.1085C} which was performed with less sophisticated models and lower spatial resolution. Our main results are:

\begin{itemize}

\item In order to take radial and vertical magnetic field components correctly into account, a 3D~disk model is vital for the simulation and interpretation of the polarized emission.

\item We find that the polarized signal which arises from aspherical grains aligned by the magnetic field does trace the magnetic field topology in both ideal observations and simulated ALMA observations. It has to be noted that the high-angular resolution of ALMA is critical for resolving the polarization pattern. Unresolved structure in the polarization pattern leads to annihilation of the signal.

\item Observations of the polarized emission of protoplanetary disk with ALMA are able to reveal the toroidal magnetic field structure. We have shown that it is even possible to observe small-scale deviations from the toroidal structure which could appear inside vortices located in the outer protoplanetary disk regions.

\item At an inclination of $45^{\circ}$ we find that small-scale structures in the magnetic field, such as a vortex, can only be observed with very high resolutions. 

\item For disks at lower inclination, we find polarization patterns corresponding to the observations of protoplanetary disks (class~I) that have been performed so far. Detailed comparisons with observations of class II disks await future detections.

\item Gaps and jumps in the dust density distributions by magnetic activity as well as vertical magnetic field components in edge-on disks are still unresolved by the polarimetric observations.

\end{itemize}

\section*{Acknowledgements}
GHMB acknowledges financial support by the DFG under contract WO857/11-1 within the frame of the DFG Priority Program1573: The Physics of the Interstellar Medium, as well as by the Millennium  Science Initiative (Chilean  Ministry  of  Economy),  through grant Nucleus RC13007. 



\bibliographystyle{mnras}
\bibliography{biblio} 



\bsp	
\label{lastpage}
\end{document}